\documentclass[reprint,prl,amsmath,amssymb,aps]{revtex4-2}

\usepackage{bbm}
\usepackage{comment}
\usepackage[normalem]{ulem}
\usepackage{siunitx}
\usepackage{enumerate}
\usepackage{graphicx}
\usepackage{dcolumn}
\usepackage{bm}
\usepackage{xcolor}
\usepackage{mathrsfs}

\usepackage{bbm}
\newcommand{\R}{{\mathbb R}}
\newcommand{\e}{{\varepsilon}}

\newcommand{\tail}{\mathcal T}
\newcommand{\head}{\mathcal H}
\newcommand{\rev}{\textcolor{black}}

\begin{document}

\preprint{APS/123-QED}

\title{Emergence and control of synchronization in networks with directed many-body interactions}

\author{Fabio Della Rossa}
\affiliation{Department of Electronics, Information, and Bioengineering, Politecnico of Milan, Italy}

\author{Davide Liuzza}
\affiliation{Department of Engineering, University of Sannio, Benevento, Italy.}

\author{Francesco Lo Iudice}
\affiliation{Department of Information Technology and Electrical Engineering, University of Naples Federico II, 80125, Naples, Italy}

\author{Pietro De Lellis}
\email{pietro.delellis@unina.it}
\affiliation{Department of Information Technology and Electrical Engineering, University of Naples Federico II, 80125, Naples, Italy}

\date{\today}

\begin{abstract}
The emergence of collective behaviors in networks of dynamical units in pairwise interaction has been explained as the effect of diffusive coupling. \rev{How does} the presence of higher-order interaction \rev{impact the onset of spontaneous or induced} synchronous behavior? Inspired by actuation and measurement constraints typical of physical and engineered systems, we propose a diffusion mechanism over hypergraphs that explains the onset of synchronization through a clarifying analogy with signed graphs. Our findings are mathematically backed by general conditions for convergence to the synchronous state.
\end{abstract}

\maketitle

A cornerstone in the literature explaining the onset of synchronized behaviour in coupled dynamical systems has been the assumption of diffusive coupling between the nodes \cite{pecora1998master,boccaletti2006complex}. Since the seminal work of Pecora and Carroll \cite{pecora1998master}, the classic equation to study the coordinated behaviour of $N$ coupled systems has been of the type
\begin{equation}
\begin{aligned}
    \dot x_i&= f(x_i,t)+\sum_{j\in\mathcal N_{\mathrm{in}}^{i}} \!\sigma_{ij}\,g(y_j-y_i),\\
    y_i&=\gamma(x_i),\hspace{14.5mm} i=1,\ldots,N,
\end{aligned}
\label{eq:classic}
\end{equation}
where $x_i\in\mathbb R^n$ and $y_i\in\R^m$ are the state and output of node $i$, $f:\R^n\times \R^+\rightarrow \R^n$ is the vector field describing the individual dynamics, $\gamma:\R^n\rightarrow\R^m$ is the output function, $g:\R^m\rightarrow \R^n$ is the coupling function, $\mathcal N_{\mathrm{in}}^i$ is the in-neighborhood of node $i$, defined as the set of nodes having an outgoing link to $i$, and $\sigma_{ij}$ is a positive scalar quantifying the coupling strength of the edge $(j,i)$.

Most of the network models in the literature can be expressed in terms of the general model \eqref{eq:classic}. For instance, \rev{it can be used to describe the dynamics in groups of identical} Kuramoto oscillators \cite{kuramoto1984}\rev{.} Furthermore, a wide range of work that detailed the mechanism underlying the synchronization of chaotic systems can be written as in \eqref{eq:classic}, as well as all works on pinning control \cite{wang2002pinning,sorrentino2007controllability,frasca2012spatial,delellis2018partial} and the classic consensus problem \cite{olfati2004consensus,bullo2020lectures}, with appropriate settings of functions $g$ and $\gamma$.

An assumption underlying the classic work on networks is the dyadic nature of the interaction among the nodes. A wide range of network systems, however, display many-body interactions that cannot be, in general, factorized in terms of pairwise interactions. This is the case of functional brain networks, where considering higher-order topological objects allowed to obtain insight on the homological structure of the brain’s functional patterns \cite{petri2014homological}. A natural framework to encode higher-order interactions is that of hypergraphs, a generalization of the concept of graph \cite{battiston2021physics,bianconi2021higher}. 

As in the case of pairwise interactions, research on synchronization in hypergraphs first focused on specific dynamics, and described how in a generalized Kuramoto model  \cite{tanaka2011multistable,skardal2019abrupt,millan2020explosive,lucas2020multiorder} the forward and backward transitions to synchronized and desynchronized states are affected by the presence of higher-order interactions. A crucial step in the study of higher-order synchronization for generic individual dynamics has been made in \cite{gambuzza2021stability}, where conditions for synchronization have been derived assuming the interaction happens on simplicial complexes, and may also be in general non-diffusive. \rev{These results have been extended in \cite{gallo2022synchronization} to a class of directed hypergraphs, denoted $M$-directed, where hyperedges appear in groups according to suitable permutations of their nodes.}

\begin{figure}
    \centering
\includegraphics[width=.8\columnwidth]{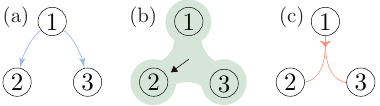}
    \caption{\rev{Different formalisms to encode directed network interactions, exemplified on the case of three nodes, where (a) is a digraph,  (b) a 1-directed hypergraph \cite{gallo2022synchronization}, and (c) the directed hypergraph we propose. Our model is the only that can capture the motivating example \eqref{eq:leader_follower}. For instance, the dynamics of node 2 would read $\dot x_2=-\sigma(x_1-x_2)$ over digraphs, whereas (b) would yield $\dot x_2=-\sigma\beta (x_1+x_3-2x_2)$ for some scalar $\beta$ according to the dynamics in \cite{gallo2022synchronization}, see Supplemental Material Section S1.}}
    \label{fig:different_couplings}
\end{figure}

\rev{The existing modeling frameworks cannot encode constraints on the feedback mechanisms that typically arise in physical and engineered network systems. For example, consider a 3-node leader-follower} consensus problem, where \rev{the leader (}node 1\rev{)} inject\rev{s} a signal to \rev{the followers (}2 and 3\rev{), but can} only \rev{measure} the\rev{ir} average state $x_{23}=(x_2+x_3)/2$\rev{. A} natural control choice would be to feed back $\sigma (x_1-x_{23})$\rev{, obtaining}
\begin{equation}\label{eq:leader_follower}
    \rev{\dot x_1=0,\qquad \dot x_2=\dot x_3=-\sigma(x_1-x_{23}),}
\end{equation}
\rev{This simple, linear three-body interaction can neither be captured by the model \eqref{eq:classic}, nor by the inherently undirected framework in \cite{gambuzza2021stability}, nor by the $M$-directed hypergraph model in \cite{gallo2022synchronization}, see Fig. \ref{fig:different_couplings}.}

\begin{figure}[t]
    \centering   \includegraphics[width=\columnwidth]{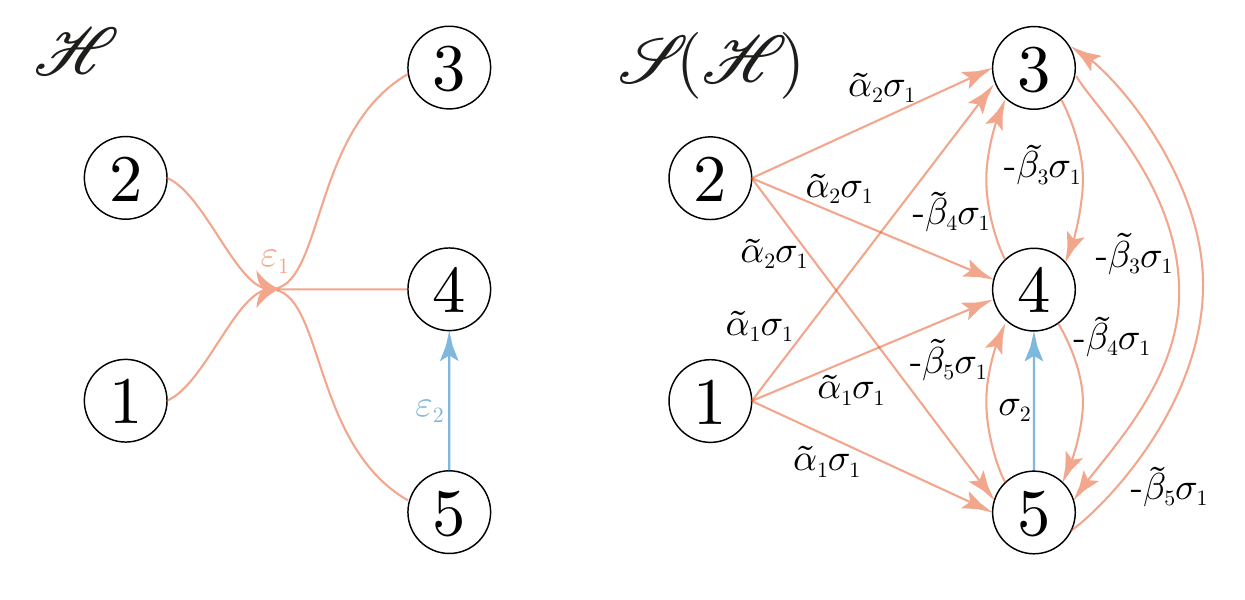}
    \caption{A sample hypergraph $\mathscr H$ and its equivalent signed graph $\mathscr S (\mathscr H)$ obtained using rule \eqref{eq:lapl_signed}.}
    \label{fig:hnet_to_net}
\end{figure}

In this Letter, we propose a novel general model of coupled dynamical systems \rev{that is able to incorporate such constraints on sensing and actuation,  with the} ambition to be the natural higher-order counterpart of the classic model \eqref{eq:classic}. 
Our model is founded on the formalism of directed hypergraph as formulated by Gallo \textit{et al.} \cite{gallo1993directed}, and the definition of hyperdiffusive coupling protocol in \cite{de2022pinning}. We consider that the interactions take place on a directed hypergraph $\mathscr H=\{\mathcal V, \mathcal E\}$, where $\mathcal V$ is the set of the $N$ nodes of the network, and $\mathcal E$ is the set of its $M$ directed hyperedges\rev{, see Fig. \ref{fig:hnet_to_net}, left panel}. A directed hyperedge $\e\in\mathcal E$ is an ordered pair $(\tail(\e),\head(\e))$ of disjoint ordered subsets of $\mathcal V$, where $\tail(\e)$ and $\head(\e)$ are the set of tails and heads of $\e$. The order of $\e$ is given by the total number of its heads and tails, and the order of the hypergraph $\mathscr H$ is the maximum order of its hyperedges.

We describe the dynamics of the $i$-th unit as
\begin{equation}\label{eq:ours}
    \begin{aligned}
        \dot x_i&= f(x_i,t)+\sum_{\e\in\mathcal E^{\cdot,i}}\sigma_{\e}g(y_{\e}^{\tau}\alpha_\e-y_{\e}^h\beta_\e),\\
        y_i&=\gamma(x_i),\hspace{22.4mm} i=1,\ldots,N,
    \end{aligned}
\end{equation}
where $\mathcal E^{\cdot,i}$ is the set of hyperedges of $\e\in\mathcal E$ such that $i\in\head(\e)$, and $\sigma_\e$ is the coupling strength associated to $\e$; $y_\e^\tau\in\R^{m\times |\tail(\e)|}$ and $y_\e ^h\in \R^{m\times|\head(\\e)|}$ are the tail and edge output matrices, obtained by juxtaposing columnwise the outputs of the nodes in $\tail(\e)$ and $\head(\e)$, respectively; $\alpha_\e$ and $\beta_\e$ are the (unit sum) vectors stacking the weights associated to the tails and heads of $\e$, respectively.

\rev{The proposed model \eqref{eq:ours} can naturally recover our motivating leader-follower example \eqref{eq:leader_follower} by considering the hypergraph in Fig. \ref{fig:different_couplings}(c), and choosing equal weights $\beta_1=\beta_2=0.5$ for the two heads (and setting $f=0$, $g(x)=\gamma(x)=x$, $\sigma_\e=1$). This simple example is paradigmatic of the wide range of diffusive interactions} of any order \rev{that the model is able to} encode\rev{, with the hyperedge tails and heads representing the nodes who inject and receive a higher-order feedback signal, respectively.} \rev{Moreover}, the hyperdiffusive coupling protocol is synchronization noninvasive \cite{gambuzza2021stability}, and therefore the synchronization manifold $x_i(t)=s(t)$ for all $i\in\mathcal V$ is invariant, with $s$ being a solution of the decoupled dynamics $\dot s=f(s,t)$. 

We now seek for an analytic tool to discriminate between the detrimental or beneficial effect higher-order interactions have on synchronizability in our general model \eqref{eq:ours}. \rev{First, we note that the argument of the nonlinear function $g$ in \eqref{eq:ours} can be rewritten as
\begin{equation}\label{eq:equiv}
\sum_{j\in\mathcal T(\e)}(\tilde\alpha_{\e})_j(y_j-y_i)-\sum_{j\in\mathcal H(\e)}(\tilde\beta_{\e})_j(y_j-y_i),
\end{equation}
Leveraging this equivalence, we can then}
define $\delta x_i=x_i - s$ as the deviation of the $i$-th node from $s$, and linearize its dynamics around $s$\rev{, thus obtaining}%
\begin{equation}\label{eq:deltax_dyn}
\dot {\delta x_i} =
 \mathrm{JF}(s) \delta x_i-\sum_{j=1}^N L_{ij} \mathrm{JG}(0) \mathrm{J\Gamma}(s) \delta x_{j},
\end{equation}
where $\mathrm{JF}\in\R^{n\times n}$, $\mathrm{JG}\in\R^{n\times m}$, and $\mathrm{J\Gamma}\in\R^{m\times n}$, are the Jacobian matrices associated to $f$, \rev{$g$}, and \rev{$\gamma$}, respectively; $L_{ij}$ is the entry $ij$ of the Laplacian matrix of the signed graph $\mathscr S(\mathscr H)$ associated to $\mathscr H$, defined as
\begin{equation}\label{eq:lapl_signed}
L_{ij}=\sum_{\e\in\mathcal E^{\cdot,\{i,j\}}}(\tilde\beta_\e)_j \sigma_\e-\sum_{\e\in\mathcal E^{j,i}}(\tilde\alpha_\e)_j \sigma_\e,
\end{equation}
where $\mathcal E^{j,i}$ is the set of hyperedges having $j$ as a tail and $i$ as a head, while $\mathcal E^{\cdot,\{i,j\}}$ is the set of hyperedges having both $i$ and $j$ as heads; $\tilde \beta_\e$ ($\tilde \alpha_e$) is a vector of $\R^N$, whose element $j$ is 0 if node $j$ is not a tail (head) of $\e$, whereas, if $j$ is a tail (head) of $\e$, it is equal to the weight associated to that tail (head). \rev{This means that the linearized dynamics of the higher-order model \eqref{eq:ours} can be equivalently represented over a directed signed graph $\mathscr{S}(\mathscr H)$ \cite{zaslavsky1982signed}, with hyperedges replaced by positive directed edges from the tails to the heads, and negative undirected edges between the heads, see Fig. \ref{fig:hnet_to_net} and Supplemental Material S2.}

\begin{figure*}[t]
    \centering   \includegraphics[width=\linewidth]{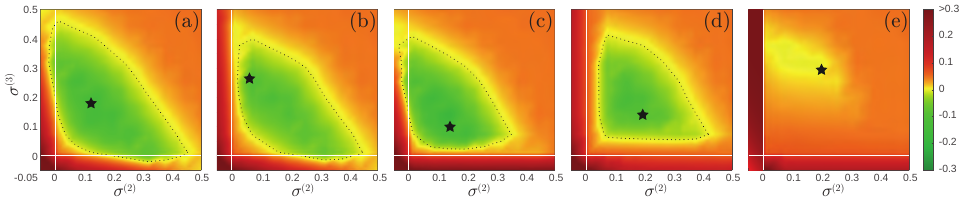}
    \caption{Synchronizability of directed hypernetworks of $N=100$ R\"ossler oscillators. Panels (a)-(e) report the colormaps of $\Lambda_{\max}$ as a function of the pair $(\sigma^{(2)},\sigma^{(3)})$ for 5 different ER hypergraphs of order 3 \rev{with $p=0.05$}, with a star identifying the pairs $(\sigma^{(2)},\sigma^{(3)})$ minimizing $\Lambda_{\max}$; when present, the black dotted curve encircles the region of the plane $(\sigma^{(2)},\sigma^{(3)})$ where $\Lambda_{\max}$ is negative and the synchronization manifold is locally asymptotically stable. \rev{The associated master stability function $\Lambda$ is reported in Supplemental Figure S3.}}
    \label{fig:Rossler}
\end{figure*}

Defining the stack vector $\delta x=[\delta x_1;\ldots;\delta x_N]$, we introduce the transformed variable $\eta=(V^{-1}\otimes I_n)\delta x$, where $V$ is \rev{the full rank matrix such that $V^{-1}LV$ is the Jordan matrix associated to $L$}, and decompose $\eta$ as the vertical stack $[\eta_1;\ldots;\eta_N]$, where $\eta_i\in\R^n$. 
Note that, regardless of the hypergraph topology, since $L$ is zero row-sum, it will always have a 0 eigenvalue (with eigenvector $\mathbbm{1}_N$) that we will denote $\lambda_1$. Therefore, \rev{from  \eqref{eq:deltax_dyn}, $\dot{\eta}_1=\mathrm{JF}(s)\eta_1$} will describe the dynamics along the synchronization manifold, irrelevant for its transversal stability. \rev{Then, to study synchronizability, we need to focus on the remaining blocks of the Jordan canonical form \cite{nishikawa2006synchronization}. A generic Jordan block of size $b$ will be associated to the transformed variables $\eta_i,\ldots,\eta_{i+b-1}$ for some $i>1$, whose dynamics are given by
\begin{subequations}\label{eq:etai}
\begin{align}
     \dot\eta_i&=\big( \mathrm{JF}(s)- \lambda_i\mathrm{JH} \big)\eta_i,\\
     \dot\eta_{i+1}&=\big( \mathrm{JF}(s)- \lambda_i\mathrm{JH} \big)\eta_{i+1}-\mathrm{JH}\eta_i,\\
     \vdots & \nonumber\\
     \dot\eta_{i+b-1}&=\big( \mathrm{JF}(s)- \lambda_i\mathrm{JH} \big)\eta_{i+b-1}-\mathrm{JH}\eta_{i+b-2},
\end{align}
\end{subequations}
where $\mathrm{JH}=\mathrm{JG}(0)\mathrm{J\Gamma}(s)$.
}

\rev{Introducing} the master equation
\begin{equation}\label{eq:master}
\dot\zeta=\Big( \mathrm{JF}(s)- \nu\mathrm{JG}(0) \mathrm{J\Gamma}(s) \Big)\zeta,
\end{equation}
where $\zeta\in\R^n$, and $\nu$ is a complex number, we can then define the master stability function (MSF) $\Lambda(\nu)$ for network \eqref{eq:ours} as the maximum Lyapunov exponent associated to \eqref{eq:master}. The stability of the synchronization manifold of network \eqref{eq:ours} will require the evaluation of $\Lambda$ at $\nu=\lambda_i$, $i=2,\ldots,N$. \rev{Indeed, i}f
\begin{equation}\label{eq:synchronizability}
\Lambda_{\max}=\max_{i=2,\ldots,N}\Lambda(\lambda_i)<0,
\end{equation}
then \rev{all Jordan blocks \eqref{eq:etai} will be asymptotically stable \cite{nishikawa2006synchronization}, and} the synchronization manifold \rev{of \eqref{eq:ours}} will be locally asymptotically stable. 

Different from the classic MSF approach on graphs, when dynamics take place over hypergraphs one needs to study $\Lambda$ also for $\nu$ with negative real-part, since the spectrum of the Laplacian matrix $L$ associated to the signed graph $\mathscr S(\mathscr H)$ may also include negative real-part eigenvalues. 

Studying the $n$-dimensional parametric master stability equation \eqref{eq:master} we derived, it is possible to gauge the impact that higher-order interactions described by model \eqref{eq:ours} have on the stability of the synchronization manifold, without the need of simulating the network dynamics. \rev{We demonstrate the potential of our approach in two paradigmatic numerical experiments on synchronization and control of network systems, where we elucidate the interplay between pairwise and high-order directed interactions and explore efficient control strategies in different directed hypergraph structures, respectively.}

\rev{{\it Synchronization.} Here, we focus on} hypergraphs of order 3 \rev{and set} all pairwise and triadic interactions \rev{to} have the same weights $\sigma^{(2)}$ and $\sigma^{(3)}$, respectively. \rev{In general, one of these f}ive scenarios may occur:
\begin{enumerate}[(a)]
    \item the synchronization manifold can be stabilized both by using only pairwise ($\sigma^{(3)}=0$) and only triadic ($\sigma^{(2)}=0$) interactions;
    \item the synchronization manifold can be stabilized by using only pairwise ($\sigma^{(3)}=0$) but not using only triadic ($\sigma^{(2)}=0$) interactions;
    \item the synchronization manifold can be stabilized by using only triadic ($\sigma^{(2)}=0$) but not using only pairwise ($\sigma^{(3)}=0$) interactions; 
    \item the synchronization manifold can only be stabilized by using both pairwise and triadic interactions; and
    \item the synchronization manifold cannot be stabilized for any value of the pair $(\sigma^{(2)},\sigma^{(3)})$.
\end{enumerate}
\rev{To investigate the prevalence of these scenarios, we randomly generated a set of 100}
Erd\"os-R\'enyi (ER) hypergraphs \rev{(with $N=100$ nodes, $p=0.05$, see Supplemental Material Sections S3).
As individual dynamics, we considered} R\"ossler \rev{chaotic} oscillators \cite{rossler1976equation}, \rev{$f(z,t)=[-z_{2}-z_{3};z_{1}+0.2 z_{2}; 0.2+z_{3}(z_{1}-7)]$,} coupling function $g$ as the identity, and output function $\gamma(z)=\allowbreak[z_1;0;0]$\rev{, so that} the MSF of the network is the same studied in  \cite{pecora1998master}, where it was first derived. \rev{For $p=0.05$, Scenarios from (a) to (e) are observed in 6, 46, 7, 40, and 1 istances, respectively, and Fig. \ref{fig:Rossler} depicts a sample occurrence of each scenario. Therefore,} in about half of the cases, \rev{the interplay between pairwise and triadic} interaction\rev{s} is key for \rev{synchronizability.}

\rev{To gauge the role of higher-order interactions for synchronizability, we consider ER hypergraphs for different values of the parameter $p$ (varied between 0.01 and 0.1 with step 0.01) modulating the expected number of hyperedges, and compare them against digraphs with the same expected cardinality $|\mathcal E |$. We observe that synchronization is more likely achieved in topologies with triadic interactions, see Supplemental Table S1. This is mainly due to the additional hyperpaths associated to hyperedges of order 3, which favor the formation of a directed spanning tree (DST) in the associated signed graph, a necessary condition to satisfy \eqref{eq:synchronizability} \cite{ahmadizadeh2017eigenvalues}. This also explains why, for increasing values of $p$, we observe a gradual transition from  Scenario (e) to (a), see Supplemental Table S2.}

\rev{{\it Leader-follower control.}}
\rev{In classic leader-follower control on digraphs, the leader can measure the output of the nodes where the control input is injected \cite{liu2008controllability}. O}ur formalism can account for a constraint \rev{on the} measurement \rev{resolution}, where the leader can only gather an aggregated measurement from groups of nodes, as typical in control applications \cite{yu2020learning,salzano2022ratiometric}. Under the tenable premise that obtaining highly resolved measurements of the aggregated state of small node groups may be more expensive or unfeasible, we \rev{evaluate how our ability to control the network varies with the resolution.}

\rev{We start with the 100 ER hypergraphs (with $p=0.05$) of $N=100$ R\"ossler systems studied above, and consider as coupling gains $(\sigma^{(2)}, \sigma^{(3)})$ those maximizing $\Lambda_{\max}$.} The leader is an \rev{extra} node, never a head of a hyperedge, \rev{so that} its dynamics \rev{$\dot x_{N+1}(t)=f(x_{N+1},t)$} are \rev{not influenced by} the rest of the nodes, \rev{and injects a feedback signal to all followers to set the synchronization manifold $s(t)$ to $x_{N+1}(t)$.}

\begin{figure}[tb]
    \centering   \includegraphics[width=\columnwidth]{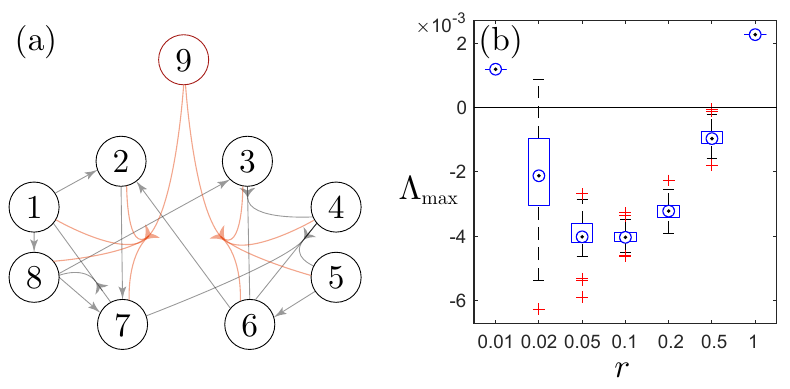}
    \caption{Panel (a) illustrates the  control strategy \rev{for $N=8$ follower nodes and a resolution $r=0.25$}: the leader \rev{(node 9)} measures the average state of two \rev{disjoint groups of} $h_c=4$ nodes. Panel (b) reports a box plot of $\Lambda_{\max}$ as a function of $r$ when controlling \rev{a sample} network of $N=100$ R\"ossler systems: \rev{any resolution $r<1$ yields lower values of $\Lambda_{\max}$ compared to $r=1$}.}
    \label{fig:control}
\end{figure}

The coarseness of the measurements \rev{the leader can take} depend on the available \emph{resolution} $\rev{r=1/h_c}$\rev{, with} $h_c$ \rev{being the number} of \rev{the} heads of the control hyperedges \rev{of which the leader measures an aggregated state.} \rev{For instance, in the $N=8$ follower nodes network in Fig.~\ref{fig:control}(a), $r=0.25$, and the leader is the unique tail of $Nr=2$ hyperedges, and can measure the aggregated state of groups of $h_c=4$ nodes. In our analysis, we vary $r$ between 0.01 (the leader measures the aggregated state of all followers) and $1$ (the standard leader-follower strategy where the leader measures the state of each follower). For intermediate values of $r$,} we consider 100 alternative selections of the control hyperedges \rev{by partitioning the followers in $Nr$ groups of cardinality $h_c$.}

\rev{We compare control performance for different resolutions in terms of the $\Lambda_{\max}$ computed on the enlarged network that includes the leader. Fig.~\ref{fig:control}(b) illustrates a paradigmatic instance for one of the 100 considered hypergraphs where, compared to standard leader-follower control ($r=1$), $\Lambda_{\max}$ decreases for any resolution $r<1$ and any selection of the control hyperedges. Note that the same result is obtained for $96$ of the $100$ considered hypergraphs, whereas for all topologies it is always possible to find resolutions $r<1$ that yield smaller $\Lambda_{\max}$ compared $r=1$.}

\rev{From condition \eqref{eq:synchronizability}, the root cause of this apparently counterintuitive phenomenon can be sought in the effect that the addition of the control hyperedges has on the spectrum of $L$. While with $r=1$ the eigenvalues are simply shifted to the right, the control hyperedges of order larger than 2 affect the shape of the spectrum, which will be enclosed by a smaller region of the complex plane, thereby facilitating control. Indeed, comparing coarser resolutions with $r=1$, a decrease of $\Lambda_{\max}$ is associated to a smaller rectangle enclosing the spectrum of $L$, except of $\lambda_1=0$ (Paerson correlation coefficient $r=0.93$, $p\text{-value}<0.001$).}

\rev{To assess how these findings generalize with network connectivity, we have analyzed the ER hypergraphs with $p$ varied between $0.01$ and $0.1$ from the synchronization study above, see Supplemental Table S3. We found that when the followers' signed network has a DST, a low resolution control can outperform the standard leader-follower control. In the absence of a DST, instead, the uncontrolled network has one or more zero eigenvalues other than $\lambda_1$, that are less likely to change with lower resolutions. This result goes beyond the specific dynamics of the R\"ossler systems, and applies to all coupled systems characterized by a MSF with a bounded stability region.}

\textit{Conclusions.} 
In this Letter, 
\rev{we} propos\rev{e} a novel and natural generalization of the classic model of diffusive interactions on digraphs to the case of multibody interactions of any order. Upon this model, we establish a powerful analogy with signed graphs, and derive a methodology to study \rev{the spontaneous or controlled emergence} of synchronization.
Through our analysis, we provide a method to discriminate whether the higher-order interaction is beneficial or detrimental for group coordination. 

\rev{We have illustrated the potential of our methodology on two relevant instances of collective behavior. In synchronization problems, our analogy with signed graphs allows to explain that higher-order interactions foster coordination, whereby they favor the formation a directed spanning tree.}
\rev{In leader-follower problems, o}ur formalism proves to be the natural way of studying and representing the measurement constraints that often appear in control of emergent behaviors. In this type of problems, we observe a nontrivial phenomenon, whereby the lack of measurement resolution not necessary \rev{hinder} network \rev{control}. \rev{On the contrary, we have shown that, when the MSF has a bounded stability region and the signed graph associated to the follower's hypergraph has a directed spanning tree, lower measurement resolutions enhance our ability to control the network.}

Our work paves the way for further studies on the interactions taking place on hypergraphs. As for the classic model \eqref{eq:classic}, also the properties of the proposed hyperdiffusive model \eqref{eq:ours} should be tested when its underlying assumptions are not met. For instance, individual differences between nodes should be properly accounted for, and different, nondiffusive types of interaction should be explored. Further, while the methodology has been demonstrated
on synthetic data, its use is envisaged in other, more detailed models of collective behavior \cite{vicsek2012collective}, as well as experimental observations on animal groups, from insect swarms to bird flocks, fish schools, and human crowds.

\begin{acknowledgments}
This work was supported by the Research Project PRIN 2017 ``Advanced Network Control of Future Smart Grids'' funded by the Italian Ministry of University and Research (2020–2023).
\end{acknowledgments}

\bibliography{IEEEabrv,biblio}
\end{document}